\documentclass[a4paper,11pt,oneside,reqno]{amsart}
\usepackage[utf8]{inputenc}
\usepackage{amsmath,amsthm,amsfonts,latexsym,amssymb,bm,enumerate}
\usepackage{ae}
\usepackage{cite}
\usepackage{lmodern}
\usepackage[T1]{fontenc}
\usepackage{slashed}

\usepackage{color}

\usepackage[colorlinks=true]{hyperref}

\usepackage[dvips]{graphicx}
\usepackage{psfrag}
\DeclareGraphicsExtensions{.eps,.art,.ART,.ps}

\usepackage{color}
\usepackage{rotating}

\newcounter{mnotecount}[section]

\newcommand{\cal}{\mathcal}

\usepackage[dvips]{graphicx}
\usepackage{psfrag}
\DeclareGraphicsExtensions{.eps,.art,.ART,.ps,.jpg}

\DeclareFontFamily{OT1}{pzc}{}
\DeclareFontShape{OT1}{pzc}{m}{it}%
              {<-> s * pzcmi8t}{}
\DeclareMathAlphabet{\mathpzc}{OT1}{pzc}%
                                {m}{it}

\newtheorem{Thm}{Theorem}[section]

\newtheorem{Lem}[Thm]{Lemma}

\newtheorem{Rmk}[Thm]{Remark}

\newcommand{\R}{\mathbb{R}}
\newcommand{\Sd}{\mathbb{S}^2}

\newcommand{\eps}{\varepsilon}

\newcommand{\uu}{\underline{u}}
\newcommand{\uv}{\underline{v}}
\newcommand{\ucc}{\underline{\cal C}}
\newcommand{\Sdois}{T}
\newcommand{\tSdois}{\tilde{T}}
\newcommand{\s}{s}
\newcommand{\calS}{\cal{F}}

\renewcommand{\leq}{\leqslant}
\renewcommand{\geq}{\geqslant}

\begin{document}

\newcounter{enumii_saved}

\begin{center}
{\Large\bf
Higher Order Linear Stability and Instability of  Reissner-Nordstr\"om's Cauchy Horizon}\\
\ \\
Jo\~ao L.\ Costa$^{\dagger\ddagger}$
 and Pedro M.\ Gir\~ao$^\ddagger$
\\

\vspace{2.2mm}
{\small $^\dagger$ISCTE - Instituto Universit\'ario de Lisboa, Lisboa, Portugal.}
\\
{\small $^\ddagger$Center for Mathematical Analysis, Geometry and Dynamical Systems,}\\
{\small Instituto Superior T\'ecnico, Universidade de Lisboa,}\\
{\small Av. Rovisco Pais, 1049-001 Lisbon, Portugal.}

\end{center}


\title{}
\author{}
\date{\today}

\begin{abstract}
We consider smooth solutions of the wave equation, on a fixed black hole region of a subextremal Reissner-Nordstr\"om (asymptotically flat, de Sitter or anti-de Sitter) spacetime, whose restrictions to the event horizon have compact support. We provide criteria, in terms of surface gravities, for the waves to remain in $C^l$, $l\geq 1$, up to and including the Cauchy horizon.
We also provide sufficient conditions for the blow up of 
solutions in $C^1$ and $H^1$. 
\end{abstract}

\maketitle

\section{Introduction}

Cauchy horizons are the spacetime boundary of the maximal Cauchy development of initial value problems for the Einstein field equations. Whenever non-empty, their existence and stability puts into question global uniqueness,
and consequently challenges the deterministic character of General Relativity. 
To understand how perturbations of a static charged black hole behave at the Cauchy horizon that lies 
in its interior, we will study solutions of the wave equation on the black hole region of fixed subextremal Reissner-Nordstr\"om 
(asymptotically flat, de Sitter or anti-de Sitter) spacetimes. 
In this framework, it is natural to consider that Cauchy horizons that allow solutions
with higher regularity are 
more stable than the ones that do not.

The stability of Cauchy horizons is a classical problem in General Relativity and, in recent years,  considerable progress has been made 
in its understanding through the mathematical analysis of wave equations.  Stability results can be found in~\cite{Anne,Sb, CF, dejan, dejan2, kehle,peterKerr,peterCauchy} and instability results 
in~\cite{luk_oh,luk_jan,DafYak,DafYak2}, and the references therein. 
For developments concerning the analysis of the full Einstein equations 
we refer to~\cite{CGNS1,CGNS2,CGNS3,CGNS4,luk_oh1,luk_oh2,daf_luk,moltel}.

Most of the literature about the linear problem focuses on stability-regularity at the $C^0$ and $H^1$ levels, in line with the modern formulations of the Strong Cosmic Censorship Conjecture. There are however some notable exceptions. In~\cite{dejan}, Gajic  provides criteria for the $C^1$ and $C^2$ extendibility of spherically symmetric waves on (asymptotically flat) extremal black holes. In the subextremal de Sitter setting, Hintz and Vasy~\cite{peterCauchy} have shown that solutions of the wave equation arising from smooth Cauchy data have $H^{1/2+\alpha/\kappa_--\epsilon}$ regularity up to the Cauchy horizon, with the degree of regularity being dictated by $\alpha$, the spectral gap 
of the operator $\Box_g$ (which also controls the decay rate of solutions along the event horizon), and 
$\kappa_-$, the Cauchy horizon's surface gravity. 
However, recent numerical computations of the spectral gap~\cite{cardoso} suggest that the regularity 
never exceeds $H^{3/2-\epsilon}$.   

Here we present criteria for higher order linear stability
of the Cauchy horizon, meaning $C^l$ with~$l\geq 1$, in a subextremal Reissner-Nordstr\"om spacetime, as well as criteria for linear instability, in both $C^1$ and $H^1$. We will achieve this by considering waves, without symmetry assumptions, whose restrictions to the event horizon have compact support. 
Although, in view of the results in~\cite{Dya,dejanDecay,holzegel}, this behavior on the event horizon 
cannot arise from generic Cauchy data, it provides a class of bona fide characteristic initial value problems for the wave equation. We will 
show that an arbitrarily high regularity at the Cauchy horizon can be obtained by increasing the order to which the wave vanishes in a direction transverse to the event horizon. 
Moreover, for this initial value problem, the role of the  
surface gravities
in determining the degree of stability of the Cauchy horizon becomes particularly transparent. 
For instance, we will prove that, as a consequence of a well known relation between surface gravities, 
if the wave only vanishes to zeroth order at the event horizon then,
in spite of having compact support on the event horizon,
it cannot be extended in $C^1$ to
any neighborhood of any point on the Cauchy horizon.
In particular, this shows that we cannot expect to obtain arbitrarily high regularity for waves up to and
including the Cauchy horizon by simply increasing their decay rate along the event horizon.

\subsection{Statement of the main results}

Let us set some basic terminology and notation.
Let $({\cal M},g)$ be a connected component of the black hole region of a subextremal Reissner-Nordstr\"om (asymptotically flat, de Sitter or anti-de Sitter) spacetime.
Denote by $\kappa_+$ and $\kappa_-$ the surface gravities of the future event horizon $\cal{H}^+$
and the future Cauchy horizon $\cal{CH}^+$, respectively, and let ${\cal H}_A^+$ and  $\cal{CH}_A^+$ denote the ``right side'' components of these horizons  (see Figure~\ref{figura}).
Let $v$ be a future increasing affine parameter of the generators of ${\cal H}_A^+$, constant on each symmetry sphere, and let $\ucc_{v_0}$ denote an ingoing null hypersurface that intersects ${\cal H}_A^+$, at $v=v_0$.
Letting $X$ be a smooth vector field which is tangent to $\ucc_{v_0}$ and transverse to ${\cal H}_A^+$, we will say that
$\phi|_{\ucc_{v_0}}$ vanishes to order~$\s\in\mathbb{Z}_0^+$ at ${\cal H}^+$
 if 
\begin{equation}
\label{order}
\phi|_{\ucc_{v_0}\cap\,{\cal H}^+}=(X\phi)|_{\ucc_{v_0}\cap\,{\cal H}^+}= \cdots = (X^\s\phi)|_{\ucc_{v_0}\cap\,{\cal H}^+}=0.
\end{equation}
We are interested in properties of functions that belong to the space
\begin{eqnarray}
\calS_\s&:=&\{\phi\in C^\infty({\cal M}\cup{\cal H}^+):\Box_g\phi=0,\nonumber\\
&&
\ \ \phi|_{\cal{H}^+\cap\{v\geq v_0\}}=0,\ \phi|_{\ucc_{v_0}}\ \mbox{vanishes to order}\ \s\ \mbox{at}\ \cal{H}^+\},
\label{Fs}
\end{eqnarray}
for a fixed $\s\in\mathbb{Z}^+_0$ and some $v_0\in\mathbb{R}$.

We may now state our four main theorems. In all of them $s$
belongs to $\mathbb{Z}^+_0$.

\begin{Thm}\label{T1}
 If $\phi\in\calS_\s$ and 
$(s+1)\kappa_+>\kappa_-$, then
 $\phi$ belongs to $C^1({\cal M}\cup\cal{CH}_A^+)$. Moreover, the
second mixed null derivatives of $\phi$ belong to $C^0({\cal M}\cup\cal{CH}_A^+)$,
the restriction of $\phi$ to symmetry spheres is $C^2$,
and $\phi$ satisfies the wave equation on the Cauchy horizon.
\end{Thm}

\begin{Thm}\label{T2} 
Let $l\geq 1$.
If $\phi\in\calS_\s$ and $(s+1)\kappa_+>l\kappa_-$, then $\phi$ belongs to $C^l({\cal M}\cup\cal{CH}_A^+)$.
\end{Thm}

\begin{Thm}\label{C1-blow-up} If the spherical mean of $\phi$ (see~\eqref{mean}) belongs to
$\calS_\s\setminus \calS_{\s+1}$ and 
$(s+1)\kappa_+<\kappa_-$, then $\phi$ does not belong to $C^1(({\cal M}\cup\cal{CH}_A^+)\cap {\cal U})$, for any open set ${\cal U}\cap \cal{CH}_A^+\neq \emptyset$.
\end{Thm}

Since the inequality $\kappa_+<\kappa_-$ is valid in the entire subextremal range of Reissner-Nordstr\"{o}m we conclude that, if $\phi\in\calS_0\setminus \calS_1$, then
it cannot be extended in $C^1$ to
any neighborhood of any point on the Cauchy horizon.

It is an easy consequence of~\cite{Sb} that if $\phi\in\calS_\s$  with
$2(s+1)\kappa_+>\kappa_-$, then $\phi$ belongs to $H^1_{{\rm loc}}(\cal{M}\cup\cal{CH}_A^+)$. We prove that this result is essentially sharp.
\begin{Thm}\label{H1-blow-up}
If the spherical mean of $\phi$ belongs to
$\calS_\s\setminus \calS_{\s+1}$ with
$2(s+1)\kappa_+<\kappa_-$, then $\phi$ does not belong to $H^1(({\cal M}\cup\cal{CH}_A^+)\cap {\cal U})$, for any open set\/\ ${\cal U}\cap \cal{CH}_A^+\neq \emptyset$.
\end{Thm}

The organization of this paper is as follows. 
In Section~\ref{setup} we explain the basic setup of our problem.
In Section~\ref{secEnergy} we recall
three energy estimates due to Sbierski. In Section~\ref{pointwise} we upgrade the previous to
pointwise estimates. In Section~\ref{t1} we prove Theorem~\ref{T1} which establishes
the existence of a classical solution up to and including the Cauchy horizon.
In Section~\ref{high} we prove Theorem~\ref{T2} concerning solutions with
higher regularity. Finally, in Section~\ref{ch} we prove
Theorems~\ref{C1-blow-up} and~\ref{H1-blow-up} about blow up in $C^1$ and in $H^1$.

\section{Setup}\label{setup}

\subsection{Some useful coordinate systems.}
\begin{figure}
\hspace*{-21mm}\includegraphics[scale=.45]{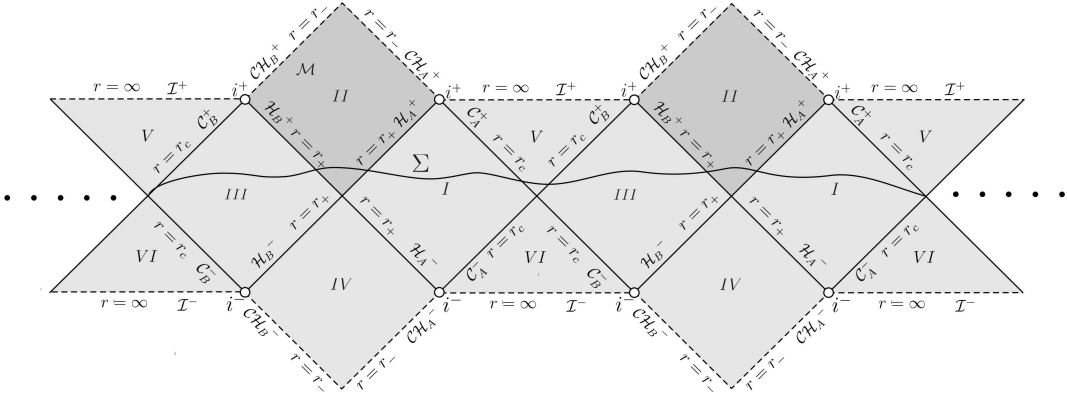}
\caption{The Penrose diagram of the Reissner-Nordstr\"{o}m-de Sitter spacetime.}
\label{figura}
\end{figure}

We will study solutions of the wave equation on a fixed background consisting
of  the black hole region of a subextremal Reissner-Nordstr\"om (asymptotically flat, de Sitter or anti-de Sitter) spacetime.
This spacetime has a
metric given in a local coordinate system by
\begin{eqnarray*}
g&=&-D\,dt^2+\frac{1}{D}\,dr^2+r^2\sigma_{\mathbb{S}^2},
\end{eqnarray*}
where
$\sigma_{\mathbb{S}^2}=d\theta^2+\sin^2\theta\,d\varphi^2$ is the round metric on the $2$-sphere, and
$$
D=D(r)=1-\,\frac{2m}{r}+\frac{e^2}{r^2}-\,\frac{\Lambda}{3}r^2.
$$
Here $m>0$ is the mass, $e\neq 0$ is the charge parameter and $\Lambda\in\mathbb{R}$ is the
cosmological constant. We will assume that
the function $D$ has at least two positive roots, the smallest of which are
$$
0<r_-<r_+.
$$
The values $r_-$ and $r_+$ correspond to the values of $r$ at the Cauchy horizon $\cal{CH}^+$ and at the event horizon ${\cal H}^+$, respectively.
The Penrose diagram of this spacetime for positive $\Lambda$ is
given in Figure~\ref{figura}.

The surface gravities of the Cauchy and event horizons, defined by
\begin{equation}
\label{surfGrav}
\kappa_-=\frac{1}{2}|D'(r_-)| \quad \text{and} \quad\kappa_+=\frac{1}{2}|D'(r_+)|,
\end{equation}
are of fundamental importance to us here. Throughout we will assume that the surface gravities do not vanish, which restricts the scope of our analysis to the subextremal setting.

For $r\in[r_-,r_+]$, we have
\begin{equation}
\label{Destimate}
 D(r)= - e^{O(1)}{(r-r_-)(r_+-r)}.
\end{equation}
Moreover, any tortoise coordinate
$$\frac{dr^*}{dr}=\frac{1}{D}$$
satisfies, for $r\in\,]r_-,r_+[$,
\begin{equation}\label{r-star}
r^*(r)=-\frac{1}{2\kappa_-}\ln|r-r_-|
+\frac{1}{2\kappa_+}\ln|r-r_+|+O(1).
\end{equation}

 The black hole region corresponds to
$$
(t,r,\omega)\in{\cal M}:=\,]-\infty,+\infty[\times]r_-,r_+[\times \Sd,
$$
a region where the function $D$ is negative, and where
$r^*$ varies in $]-\infty,+\infty[$.

We will often rely on the double null coordinates $(\uu,\uv)\in\,]-\infty,+\infty[^2$ given in
terms of $t$ and $r$ by
$$
\begin{cases}
\uu=t-r^*(r),\\
\uv=t+r^*(r).
\end{cases}
$$
In these coordinates the metric takes the form
\begin{eqnarray*}
g
&=&-D\,d\uu d\uv+r^2\sigma_{\mathbb{S}^2}.
\end{eqnarray*}
Clearly, we have
\begin{equation}\label{r-star-vu}
r^*(r)=\frac{\uv-\uu}{2}.
\end{equation}
Note that the event horizon corresponds
to $\uu=+\infty$ and the Cauchy horizon corresponds to $\uv=+\infty$.
Since our double null
coordinates are singular at these horizons,
at the event horizon we change from $(\uu,\uv,\omega)$ coordinates to $(v,r,\omega)$ coordinates
using
$$
\begin{cases}
\uu=v-2r^*(r),\\
\uv=v,
\end{cases}\qquad
\begin{cases}
\partial_{\uu}=-\,\frac{D}{2}\partial_r,\\
\partial_{\uv}=\partial_v+\frac{D}{2}\partial_r.
\end{cases}
$$
In these coordinates the metric becomes
$$
g=-D\,dv^2+2\,dvdr+r^2\sigma_{\mathbb{S}^2}.
$$
At the Cauchy horizon we change from $(\uu,\uv,\omega)$ coordinates to $(u,\tilde{r},\omega)$
coordinates using
$$
\begin{cases}
\uu=u,\\
\uv=u+2r^*(\tilde{r}),
\end{cases}\qquad
\begin{cases}
\partial_{\uu}=\partial_u-\,\frac{D}{2}\partial_{\tilde{r}},\\
\partial_{\uv}=\frac{D}{2}\partial_{\tilde{r}}.
\end{cases}
$$
In these coordinates the metric is written as
$$
g=-D\,du^2-2\,dud\tilde{r}+r^2\sigma_{\mathbb{S}^2}.
$$
Note that to change from $(v,r,\omega)$ coordinates to $(u,\tilde{r},\omega)$ coordinates we can use
$$
\begin{cases}
u=v-2r^*(r),\\
\tilde{r}=r,
\end{cases}
\qquad
\begin{cases}
\partial_u=\partial_v,\\
\partial_{\tilde{r}}=\frac{2}{D}\partial_v+\partial_r.
\end{cases}
$$
By abuse of notation, we will write $\phi(\uu,\uv,\omega)=\phi(v,r,\omega)=
\phi(u,r,\omega)$.

It is important to note that the vector field $\partial_t=\partial_v
=\partial_u$ is Killing. We also denote by ${\it \Omega}_i$, for $i=1,2,3$, the generators of spherical symmetry, and
just by ${\it \Omega}$ any one of the three.
The vector fields ${\it \Omega}$
are also Killing.

\subsection{The wave equation}
Define
\begin{equation}
\label{Gn}
G_n=\frac{1}{r}D+\frac{n}{2}D',
\end{equation}
\begin{equation}
\label{S}
S=-\,\frac{1}{r}\partial_v-\frac{1}{2}\slashed\Delta
\qquad
\mbox{and}
\qquad
\tilde{S}=\frac{1}{r}\partial_u-\frac{1}{2}\slashed\Delta,
\end{equation}
with $\slashed\Delta\phi$ denoting the spherical laplacian of $\phi$,
$$
\slashed\Delta\phi=\frac{1}{r^2}\partial_\theta^2\phi+\frac{1}{r^2}\cot\theta\,\partial_\theta\phi
+\frac{1}{r^2\sin^2\theta}\partial_\varphi^2\phi.
$$
The wave equation,
$$
\Box_g\phi=\nabla^\mu\nabla_\mu\phi=0,
$$
is equivalent to both
\begin{equation}\label{wave_v}
\partial_{\uv}\partial_r\phi+G_1\partial_r\phi=S(\phi)
\end{equation}
and
\begin{equation}\label{wave_u}
\partial_{\uu}\partial_{\tilde r}\phi-G_1\partial_{\tilde r}\phi=-\tilde{S}(\phi).
\end{equation}

\subsection{The energy-momentum tensor}
Recall that to a scalar function $\phi$ we may associate the energy-momentum tensor
$$
T_{\mu\nu}=\partial_\mu\phi\partial_\nu\phi-\frac{1}{2}
g_{\mu\nu}(\partial_\alpha\phi\,\partial^\alpha\phi),
$$
whose relevance for the study of solutions of the wave equation
stems from the fact that its divergence satisfies
$$
\nabla^\mu T_{\mu\nu}=(\Box_g\phi)\,
\partial_\nu\phi.
$$
Our energy estimates for $\phi$ will be obtained by applying the Divergence Theorem
to certain currents, which are contractions of the energy-momentum tensor
with appropriate vector fields. It will be useful to have the
expression of the energy-momentum tensor in coordinates.
One readily checks that
$$
\begin{cases}
T(\partial_{\uu},\partial_{\uu})=(\partial_{\uu}\phi)^2,\\
T(\partial_{\uu},\partial_{\uv})=\frac{D}{4}|\slashed\nabla\phi|^2,\\
T(\partial_{\uv},\partial_{\uv})=(\partial_{\uv}\phi)^2.
\end{cases}
$$
Again, $\slashed\nabla\phi$ denotes the spherical gradient of $\phi$,
$$
\slashed\nabla\phi=\frac{1}{r^2}(\partial_\theta\phi)\,\partial_\theta+\frac{1}{r^2\sin^2\theta}
(\partial_\varphi\phi)\,\partial_\varphi,
$$
and
$$
|\slashed\nabla\phi|^2=r^2\sigma_{\mathbb{S}^2}(\slashed\nabla\phi,\slashed\nabla\phi).
$$

\subsection{Energy identities and the Divergence Theorem}
We will apply the Divergence Theorem in regions bounded
by hypersurfaces $\ucc_v$, where $\uv$ is constant equal to $v$, hypersurfaces ${\cal C}_u$,
where $\uu$ is constant equal to $u$, and hypersurfaces $\Sigma_r$, where
the geometric variable~$r$ is constant
equal to $r$.
Denoting by $n_{\ucc_v}$, $n_{{\cal C}_u}$ and $n_{\Sigma_r}$ the corresponding
normals, with
$n_{\Sigma_r}$ unit and all three future directed, and denoting by
$dV_{\ucc_v}$, $dV_{{\cal C}_u}$ and $dV_{\Sigma_r}$ the corresponding volume elements,
we have
$$
\begin{array}{rclcrcl}
n_{\ucc_v}&=&-\partial_r,&\qquad&dV_{{\ucc_v}}&=&r^2\,drd\omega,
\\
\vspace{-3mm}
\\
n_{{\cal C}_u}&=&-\partial_{\tilde r},&\qquad&dV_{{\cal C}_u}&=&r^2\,d\tilde{r}d\omega,
\\
\vspace{-3mm}
\\
n_{\Sigma_r}&=&\frac{-\partial_u+D\partial_{\tilde r}}{\sqrt{-D}},&\qquad& dV_{\Sigma_r}&=&\sqrt{-D}\,r^2\,dud\omega,
\end{array}
$$
where $d\omega$ is the volume form associated to $\sigma_{\mathbb{S}^2}$. Note that along the null hypersurfaces there is no natural choice of normal or volume form, so one can just
choose a convenient normal and then
let the Divergence Theorem determine the volume form.

Our currents will be vector fields of the form
$$
J^N_\nu=J^N_\nu(\phi):=T_{\mu\nu}N^\mu=T_{\mu\nu}(\phi)N^\mu,
$$
with $N$ timelike and future pointing, so that if $\phi$ is a solution of
the wave equation, then
\begin{eqnarray*}
\nabla^\nu J^N_\nu&=&
(\nabla^\nu T_{\mu\nu})N^\mu+T_{\mu\nu}\nabla^\nu N^\mu\\ &=&
(\Box_g\phi)\,N\cdot\phi+T_{\mu\nu}\nabla^\nu N^\mu=T_{\mu\nu}\nabla^\nu N^\mu.
\end{eqnarray*}
Our choices of $N$ will be such that $T_{\mu\nu}\nabla^\nu N^\mu$ is nonnegative.
We denote by
\begin{eqnarray*}
&{\cal C}_u(r_2,r_1)={\cal C}_u\cap\{r_2\leq r\leq r_1\},&\\
&{\ucc}_v(r_2,r_1)={\ucc}_v\cap\{r_2\leq r\leq r_1\}.&
\end{eqnarray*}
Applying the Divergence Theorem to the current $J^N_\nu$ in the region
$${\cal R}=\{u\geq u_0\wedge v\leq v_0\wedge r_2\leq r\leq r_1\}$$
\begin{figure}
\begin{turn}{45}
\includegraphics[scale=.8]{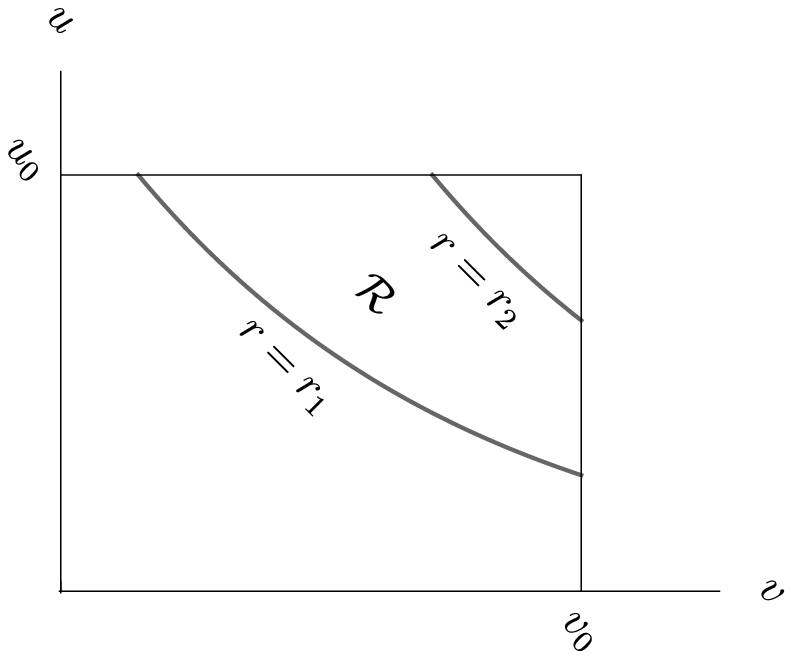}
\end{turn}

\vspace{-11mm}

\caption{The region $\cal{R}$.}
\label{fig}
\end{figure}
(see Figure~\ref{fig}) we get the energy identity
\begin{eqnarray*}
\int\displaylimits_{\Sigma_{r_1}\cap\{u\geq u_0 \wedge v\leq v_0\}}J^N_\nu n^\nu_{\Sigma_{r_1}}\,dV_{\Sigma_{r_1}}
-\int\displaylimits_{\ucc_{v_0}(r_2,r_1)}J^N_\nu n^\nu_{\ucc_{v_0}}\,dV_{\ucc_{v_0}}\qquad\qquad&&\\
-\int\displaylimits_{\cal{C}_{u_0}(r_2,r_1)}J^N_\nu n^\nu_{\cal{C}_{u_0}}\,dV_{\cal{C}_{u_0}}
-\int\displaylimits_{\Sigma_{r_2}\cap\{u\geq u_0 \wedge v\leq v_0\}}J^N_\nu n^\nu_{\Sigma_{r_2}}\,dV_{\Sigma_{r_2}}\qquad\qquad&&\\
=\iint\displaylimits_{\cal{R}} \nabla^\nu J^N_\nu\,dV_{\cal{M}}.&&
\end{eqnarray*}

For a hypersurface~$S$, the integral
$$
\int_SJ^N_\nu n^\nu_S\,dV_S=\int_ST(N,n_S)\,dV_S
$$
controls first order derivatives of $\phi$. Let us give an example by defining, near the Cauchy horizon, $N=\tilde{N}_b=-\partial_{\uu}-\partial_{\tilde r}$. This choice leads to
\begin{eqnarray*}
T(\tilde{N}_b,n_{\ucc_v})\,dV_{\ucc_v}&=&\left(-\,\frac{2}{D}T(\partial_{\uu},\partial_{\uu})-\,
\frac{4}{D^2}T(\partial_{\uu},\partial_{\uv})\right)r^2\,drd\omega\\
&=&\left[-\,
\frac{1}{D}\left(2(\partial_{\uu}\phi)^2+|\slashed\nabla\phi|^2
\right)\right]r^2\,drd\omega,
\end{eqnarray*}
\begin{eqnarray}
T(\tilde{N}_b,n_{{\cal C}_u})\,dV_{{\cal C}_u}&=&\left(\frac{2}{D}T(\partial_{\uu},\partial_{\uv})+\frac{4}{D^2}
T(\partial_{\uv},\partial_{\uv})\right)r^2\,d\tilde{r}d\omega\nonumber\\
&=&\left[(\partial_{\tilde r}\phi)^2+\frac{1}{2}|\slashed\nabla\phi|^2
\right]r^2\,d\tilde{r}d\omega,\nonumber
\end{eqnarray}
\begin{eqnarray*}
&&T(\tilde{N}_b,n_{\Sigma_r})\,dV_{\Sigma_r}\\
&&\qquad\qquad=\left(T(\partial_{\uu},\partial_{\uu})
+\left(\frac{2}{D}-1\right)T(\partial_{\uu},\partial_{\uv})
-\,\frac{2}{D}T(\partial_{\uv},\partial_{\uv})\right)r^2\,dud\omega\nonumber\\
&&\qquad\qquad=\left[(\partial_{\uu}\phi)^2-\,\frac{D}{2}(\partial_{\tilde r}\phi)^2+\frac{1}{2}
\left(1-\,\frac{D}{2}\right)|\slashed\nabla\phi|^2
\right]r^2\,dud\omega.
\nonumber
\end{eqnarray*}
Note that the expressions inside the square parentheses above are
nonnegative as required by the fact that
energy-momentum tensor satisfies the Dominant Energy Condition.

For $-1\leq D\leq 0$ we have
\begin{eqnarray*}
T(\tilde{N}_b,n_{\Sigma_r})&\leq&\left(2(\partial_u\phi)^2-D(\partial_{\tilde r}\phi)^2+\frac{3}{4}|\slashed\nabla\phi|^2
\right)\frac{1}{\sqrt{-D}}
\nonumber
\end{eqnarray*}
and
\begin{eqnarray*}
T(\tilde{N}_b,n_{\Sigma_r})&\geq&\left(\frac{1}{2}(\partial_u\phi)^2-\,
\frac{D}{4}(\partial_{\tilde r}\phi)^2+\frac{1}{2}|\slashed\nabla\phi|^2
\right)\frac{1}{\sqrt{-D}},
\nonumber
\end{eqnarray*}
so that for $r$ sufficiently close to $r_-$ we have
\begin{equation}
T(\tilde{N}_b,n_{\ucc_v})\geq 0,\label{blue-v}
\end{equation}
\begin{equation}
T(\tilde{N}_b,n_{{\cal C}_u})\sim (\partial_{\tilde r}\phi)^2+|\slashed\nabla\phi|^2,
\label{blue-u}
\end{equation}
\begin{equation}
T(\tilde{N}_b,n_{\Sigma_r})\sim\left((\partial_u\phi)^2-D(\partial_{\tilde r}\phi)^2+|\slashed\nabla\phi|^2
\right)\frac{1}{\sqrt{-D}}.
\label{blue-r}
\end{equation}
The notation $f\sim h$ means that there exist positive constants $c$ and $C$ such that
$cf\leq h\leq Cf$.
The blue-shift vector field, which will play a relevant role below, satisfies
$$N_b=-(1+O(r-r_-))(\partial_{\uu}+\partial_{\tilde r})$$ and so expressions~\eqref{blue-v},
\eqref{blue-u} and~\eqref{blue-r} are also valid if we replace $\tilde{N}_b$ by ${N}_b$.

\section{Basic energy estimates}
\label{secEnergy}
We denote by
\begin{eqnarray*}
&\Sigma_r(u_0)=\Sigma_r\cap\{u\geq u_0\},	\quad
\Sigma_r(v_0)=\Sigma_r\cap\{v\geq v_0\}.&
\end{eqnarray*}
We will now recall some basic energy estimates. The first one applies to the red-shift region.
According to~\cite[p.\ 113, (4.5.5)]{Sb} we have
\begin{Lem}\label{u-decay}
For every $\delta>0$,
there exists a future directed timelike time invariant vector field $N_r$ in
${\cal M}\cap {\cal H}^+$, a $r_-<r_0<r_+$ and
a constant $C>0$ such that we have
\begin{equation}\label{S1}
\int_{\Sigma_{r_0}(u)}J^{N_r}(\phi)\cdot n_{\Sigma_{r_0}}\leq Ce^{-(2\s+2-\delta)\kappa_+u},
\end{equation}
for all $\phi$ belonging to $\calS_\s$.
\end{Lem}
At the event horizon $N_r=\partial_{\uv}-\partial_r$ and $N_r$ satisfies
$T_{\mu\nu}\nabla^\nu N_r^\mu\geq 0$ for $r_0\leq r\leq r_+$.

The second energy estimate applies to the no-shift region.
According to the proof of~\cite[Lemma~4.5.6]{Sb} we have
\begin{Lem}
	Given $r_-<r_1<r_0<r_+$ and a future directed timelike time invariant
	vector field $N$ in ${\cal M}$, there exists a constant $C>0$ such that
\begin{equation}\label{S2}
\int_{{\cal C}_{u}(r_1,r_0)}J^{N}(\phi)\cdot n_{{\cal C}_{u}}+
\int_{\Sigma_{r_1}(u)}J^{N}(\phi)\cdot n_{\Sigma_{r_1}}\leq C
\int_{\Sigma_{r_0}(u)}J^{N}(\phi)\cdot n_{\Sigma_{r_0}},
\end{equation}	
for all $\phi\in C^\infty({\cal M})$ satisfying $\Box_g\phi=0$.
\end{Lem}
Note that, in the previous case, $T_{\mu\nu}\nabla^\nu N^\mu$ might be negative but we can apply the Divergence
Theorem with vector field $\tilde{N}=e^{C(r_0,r_1)r}N$ and $C$ sufficiently large so that
$T_{\mu\nu}\nabla^\nu \tilde{N}^\mu$ is nonnegative.

The third energy estimate applies to the blue-shift region. According to the proof of~\cite[Proposition~4.5.8]{Sb} we have
\begin{Thm}\label{Sb}
Assume $2(\s+1)\kappa_+>\kappa_->0$.
For every sufficiently small~$\delta$, $\eps>0$,
there exists a future directed timelike time invariant vector field $N_b$ in
${\cal M}\cup {\cal C \cal H}^+$, a $r_1>r_-$ and a constant $C>0$
such that, for all $r_1>r_2>r_-$,
we have\/ {\rm (see \cite[p.\ 114, last line]{Sb})}
$$
\int_{\Sigma_{r_1}(u_0)}e^{(2\s+2-\delta)\kappa_+u}J^{N_b}(\phi)\cdot n_{\Sigma_{r_1}}\leq C
$$
and\/ {\rm (see \cite[p.\ 117, (4.5.10)]{Sb} together with the previous inequality)}
\begin{eqnarray}
&&e^{\kappa_-(1+\eps)u_0}\int_{{\cal C}_{u_0}(r_2,r_1)}J^{N_b}(\phi)\cdot n_{{\cal C}_{u_0}}+
\int_{\Sigma_{r_2}(u_0)}
e^{\kappa_-(1+\eps)u}J^{N_b}(\phi)\cdot n_{\Sigma_{r_2}}\nonumber\\
&&\qquad\qquad\qquad\leq \int_{\Sigma_{r_1}(u_0)}e^{\kappa_-(1+\eps)u}J^{N_b}(\phi)\cdot n_{\Sigma_{r_1}}
\nonumber\\
&&\qquad\qquad\qquad\leq Ce^{-[(2\s+2-\delta)\kappa_+-\kappa_-(1+\eps)]u_0},\label{first-inequality}
\end{eqnarray}
for all $\phi$ belonging to $\calS_\s$.

Moreover, the function $\phi$ belongs to
$H^1_{{\rm loc}}(\cal{M}\cup\cal{CH}_A^+)$.
\end{Thm}
At the Cauchy horizon $N_b=-\partial_{\uu}-\partial_{\tilde r}$ and $N_b$ satisfies
$T_{\mu\nu}\nabla^\nu N_b^\mu\geq 0$ for $r_-\leq r\leq r_1$.

\section{Pointwise estimates for $\partial_u\phi$, $\slashed\nabla\phi$ and $\slashed\Delta\phi$}
\label{pointwise}

Let $\phi\in\calS_\s$. Throughout this section we will assume that $2(s+1)\kappa_+>\kappa_->0$.
We will fix $\kappa$ satisfying $\kappa<(\s+1)\kappa_+$.
The objective of the next three subsections is to prove that the three estimates
\begin{eqnarray}
|\partial_u\phi|(u,r,\,\cdot\,)&\leq& Ce^{-\kappa u},\label{one}\\
|\slashed\nabla\phi|(u,r,\,\cdot\,)&\leq& Ce^{-\kappa u},\label{two}\\
|\slashed\Delta\phi|(u,r,\,\cdot\,)&\leq& Ce^{-\kappa u}\label{three}
\end{eqnarray}
hold for $r_-\leq r\leq r_0<r_+$ and $u\in\mathbb{R}$.
We choose a $\delta>0$ satisfying $\kappa\leq (\s+1-\delta)\kappa_+$.

\subsection{Estimates for $r=r_0$}
\label{subSecR0}

In a parallel manner to~\eqref{blue-r}, there exists $r_0<r_+$
such that
\begin{equation}\label{close}
J^{N_r}_\mu(\phi)n_{\Sigma_r}^\mu\,dV_{\Sigma_r}\sim\left((\partial_v\phi)^2-D(\partial_r\phi)^2+
|\slashed\nabla\phi|^2\right)r^2\,dvd\omega
\end{equation}
holds for $r_0\leq r<r_+$.

To obtain a uniform bound on $\partial_u\phi$ we use the following five ingredients:

{\bf (i)}
From~\eqref{S1} and the fact that $u=v-2r^*(r)$ (or from~\cite[p.\ 113, (4.5.5)]{Sb}),
for $v\geq v_0$,
we have that
\begin{equation}\label{estimate}
\int_{\Sigma_{r_0}(v)}
J^{N_r}(\phi)\cdot n_{\Sigma_{r_0}}\leq
Ce^{-(2\s+2-\delta)\kappa_+v},
\end{equation}
for a red-shift vector field  that satisfies
$N_r=(1+O(r_+-r))(\partial_v-\partial_r)$.
Since $\partial_v$ and the vector fields ${\it \Omega}$ are Killing, and $\phi\in\calS_\s$ implies that
$\partial^l_v\Omega^I\phi\in\calS_\s$, for every $l\in \mathbb{Z}^+_0$ and every multi-index $I$
(see Remark~\ref{dv}),
we see that estimate~\eqref{estimate} holds with $\phi$ replaced by $\partial^l_v\Omega^I\phi$.

{\bf (ii)} We now apply Sobolev's inequality in symmetry spheres and~\eqref{close} to obtain
\begin{eqnarray}
	&&c\int_{v_1}^{v_2}\sup_{\omega\in\Sd}(\partial_v\phi)^2(v,r_0,\omega)\,dv\leq
	\int_{v_1}^{v_2}\|\partial_v^2\phi(v,r_0,\,\cdot\,)\|^2_{H^2(\Sd)}\,dv
    \nonumber
    \\
	&&\qquad\qquad\leq C\int_{\Sigma_{r_0}(v_1)}
	J^{N_r}_\mu(\partial_v\phi)n_{\Sigma_{r_0}}^\mu\,dV_{\Sigma_{r_0}}
  \nonumber
    \\
	&&\qquad\qquad\ \ \,+C\sum_{i=1}^3\int_{\Sigma_{r_0}(v_1)}
	J^{N_r}_\mu({\it \Omega}_i\partial_v\phi)n_{\Sigma_{r_0}}^\mu\,dV_{\Sigma_{r_0}}
  \nonumber
    \\
	&&\qquad\qquad\ \ \,+C\sum_{i=1}^3\int_{\Sigma_{r_0}(v_1)}
	J^{N_r}_\mu({\it \Omega}_i^2\partial_v\phi)n_{\Sigma_{r_0}}^\mu\,dV_{\Sigma_{r_0}}
  \nonumber
\end{eqnarray}
and then we use~\eqref{estimate} to conclude that
\begin{equation}
\label{sobolevSphere}
\int_{v_1}^{v_2}\sup_{\omega\in\Sd}(\partial_v\phi)^2(v,r_0,\omega)\,dv\leq Ce^{-(2\s+2-\delta)\kappa_+v_1}.
\end{equation}
{\bf (iii)} We recall~\cite[Lemma~4.5]{CF}.
\begin{Lem}
\label{lemmaExp}
	Let $f:[t_0,\infty[\to\R$ and assume that for some $\alpha_1, C>0$, and for all
	$t\geq t_0$,
	$$
	\int_t^\infty f(s)\,ds\leq Ce^{-\alpha_1 t}.
	$$
	Then, for all $0<\alpha_2<\alpha_1$,
	$$
	\int_t^\infty e^{\alpha_2s} f(s)\,ds\leq Ce^{-(\alpha_1-\alpha_2)t}.
	$$
\end{Lem}

Below we will take $\alpha_1=(2\s+2-\delta)\kappa_+$ and $\alpha_2=(2\s+2-2\delta)\kappa_+$.

{\bf (iv)} If we take squares of both sides of
\begin{eqnarray*}
\partial_v\phi(v_2,r_0,\omega)=\int_{v_1}^{v_2}\partial^2_v\phi(v,r_0,\omega)dv + \partial_v\phi(v_1,r_0,\omega)
\end{eqnarray*}
and then apply H\"older's inequality we get
\begin{eqnarray*}
(\partial_v\phi)^2(v_2,r_0,\omega)
&\leq&
2 \left(\int_{v_1}^{v_2}\partial^2_v\phi(v,r_0,\omega)dv\right)^2 + 2(\partial_v\phi)^2(v_1,r_0,\omega)
\\
&\leq&
C e^{-(2\s+2-2\delta)\kappa_+ v_1} \int_{v_1}^{v_2}e^{(2\s+2-2\delta)\kappa_+ v}(\partial^2_v\phi)^2(v,r_0,\omega)dv
\\
&&\,+
2(\partial_v\phi)^2(v_1,r_0,\omega).
\end{eqnarray*}

We can then use~\eqref{sobolevSphere} and Lemma~\ref{lemmaExp} to conclude that
\begin{eqnarray}
\label{supEstimate}
&&\sup_{\omega\in\Sd}(\partial_v\phi)^2(v_2,r_0,\omega)
\nonumber
\\
&&\qquad\qquad\leq C
e^{-(2\s+2-2\delta)\kappa_+v_1}\int_{v_1}^{v_2}e^{(2\s+2-2\delta)\kappa_+v}\|\partial_v^2\phi(v,r_0,\,\cdot\,)\|^2_{H^2(\Sd)}\,dv
\nonumber
\\
&&\qquad\qquad\ \ \ \,+
2\sup_{\omega\in\Sd}(\partial_v\phi)^2(v_1,r_0,\omega)
\nonumber
\\
&&\qquad\qquad\leq C
e^{-(2\s+2-2\delta)\kappa_+v_1}+2\sup_{\omega\in\Sd}(\partial_v\phi)^2(v_1,r_0,\omega).
\end{eqnarray}

{\bf (v)} From~\cite[Lemma~4.2]{CF}, we know that
\begin{Lem}\label{decay}
Let $t_0\geq 0$, and $b, B, M, \Delta>0$. Let $f:[t_0,\infty[\to\R^+$ be continuous
and such that
$$
f(t_2)+b\int_{t_1}^{t_2}f(t)\,dt\leq Mf(t_1)+Be^{-\Delta t_1},
$$
for all $t_0\leq t_1\leq t_2$. Then, for $\alpha<\min\bigl\{\frac{b}{M},\Delta\bigr\}$,
we have
$$
f(t)\leq C_\alpha(1+f(t_0))e^{-\alpha t},
$$
for all $t\geq t_0$.
\end{Lem}

If we consider the function $f(v)=\sup_{\omega\in\Sd}(\partial_v\phi)^2(v,r_0,\omega)$ and add a large multiple of $\int_{v_1}^{v_2} f(v)dv$ (which satisfies~\eqref{sobolevSphere}) to both sides of~\eqref{supEstimate},
we can apply the previous lemma
to obtain the pointwise estimate
$$
|\partial_v\phi|(v,r_0,\,\cdot\,)\leq Ce^{-\kappa v},
$$
for $\kappa<(\s+1-\delta)\kappa_+$ and $v\geq v_0$.

Note that the constant $C$, in the last estimate, is uniform in $\omega$ because
$\partial_v\phi(v_0,r_0,\,\cdot\,)$ is a bounded function of $\omega$.
For $r=r_0$ we also have~\eqref{one}
since $\partial_u=\partial_v$ and $v=u+2r^*(r)$.

Since estimate~\eqref{estimate} also holds with $\phi$ replaced by ${\it \Omega}^I\phi$, for any multi-index $I$,
we can repeat the
proceeding of the previous paragraphs
to obtain~\eqref{two} and~\eqref{three}, at $r=r_0$,
with constants $C$, once again,  uniform in $\omega$.

\subsection{Estimates for $r_1\leq r\leq r_0$}

Let $N$ be a timelike future directed and time independent vector field. Similarly to~\eqref{blue-u}, we have
$$
J^N_\mu(\phi)n_{{\cal C}_u}^\mu\,dV_{{\cal C}_u}
\sim \left((\partial_{\tilde r}\phi)^2+|\slashed\nabla\phi|^2\right)r^2\,drd\omega.
$$
Since $\partial_u=\partial_v$ is Killing, and $\phi\in\calS_\s$ implies that
$\partial_u\phi\in\calS_\s$, ${\it \Omega}\partial_u\phi\in\calS_\s$ and
${\it \Omega}^2\partial_u\phi\in\calS_\s$, for $r_1\leq r\leq r_0$, by~\eqref{S2} we have
\begin{eqnarray*}
	&&\int_{r}^{r_0}\|\partial_{\tilde r}\partial_u\phi(u,\tilde r,\,\cdot\,)\|^2_{H^2(\Sd)}\,d\tilde r\\
	&&\qquad\leq C\int_{{\cal C}_u(r,r_0)}
	J^N_\mu(\partial_u\phi)n_{{\cal C}_u}^\mu\,dV_{{\cal C}_u}+C\sum_{i=1}^3\int_{{\cal C}_u(r,r_0)}
	J^N_\mu({\it \Omega}_i\partial_u\phi)n_{{\cal C}_u}^\mu\,dV_{{\cal C}_u}\\
	&&\qquad\ \ \,+C\sum_{i=1}^3\int_{{\cal C}_u(r,r_0)}
	J^N_\mu({\it \Omega}_i^2\partial_u\phi)n_{{\cal C}_u}^\mu\,dV_{{\cal C}_u}\\
	&&\qquad\leq Ce^{-(2\s+2-\delta)\kappa_+u}.
\end{eqnarray*}
This together with~\eqref{one}, applied with $r=r_0$, implies that
\begin{eqnarray*}
	&&\sup_{\omega\in\Sd}(\partial_u\phi)^2(u,r,\omega)\\
	&&\qquad\qquad\leq C
	\int_{r}^{r_0}\|\partial_{\tilde r}\partial_u\phi(u,\tilde r,\,\cdot\,)\|^2_{H^2(\Sd)}\,d\tilde r+
	2\sup_{\omega\in\Sd}(\partial_u\phi)^2(u,r_0,\omega)\\
	&&\qquad\qquad\leq Ce^{-(2\s+2-2\delta)\kappa_+u}.
\end{eqnarray*}
Thus, we obtain~\eqref{one}
for $r_1\leq r\leq r_0$.

Applying~\eqref{S2} to ${\it\Omega}\phi$, ${\it\Omega}^2\phi$, ${\it\Omega}^3\phi$ and ${\it\Omega}^4\phi$, and using~\eqref{two} and~\eqref{three} for $r=r_0$,
we obtain~\eqref{two} and~\eqref{three} for $r_1\leq r\leq r_0$.
The constants $C$ do not depend on $r\in[r_1,r_0]$ or $\omega$.

\subsection{Estimates for $r_-<r\leq r_1$}

In this region, according to~\eqref{first-inequality}, we have
\begin{equation}
\int_{{\cal C}_{u}(r_1,r_-)}J^{N_b}(\phi)\cdot n_{{\cal C}_{u}}\leq Ce^{-(2\s+2-\delta)\kappa_+u}.
\label{estimate-one}
\end{equation}
Applying~\eqref{estimate-one} to $\phi$,
${\it\Omega}\phi$ and ${\it\Omega}^2\phi$ implies that $\phi$ extends continuously to ${\cal C}{\cal H}^+$ along segments of constant $u$. Moreover, $\phi(\,\cdot\,,r_-,\,\cdot\,)$
is the uniform  limit of $\phi(\,\cdot\,,r,\,\cdot\,)$ as $r\searrow r_-$ (for $(u,\omega)\in[\overline{u},\overline{U}]\times\Sd$, where $-\infty<\overline{u}<\overline{U}<\infty$). Arguing as in~\cite[Proposition~5.2, Step~2]{CGNS3}, $\phi$
is continuous in ${\cal M}\cup {\cal C}{\cal H}^+$.
The same reasoning can be used to show that $\partial_u\phi$, ${\it\Omega}\phi$
and ${\it\Omega}^2\phi$
extend continuously to~${\cal C}{\cal H}^+$. A simple argument implies that
the derivative of the continuous extension of $\phi$ with respect to $u$ exists and
coincides with the continuous extension of $\partial_u\phi$. Analogous statements apply
to ${\it\Omega}\phi$ and ${\it\Omega}^2\phi$.
Using~\eqref{estimate-one}, and reasoning as we did in the region $r_1\leq r\leq r_0$,
we see that
\eqref{one}, \eqref{two} and~\eqref{three} hold
for $r_-<r\leq r_1$.

\section{Existence of a classical solution up to the Cauchy horizon}\label{t1}

Henceforth, by ``up to the Cauchy horizon'' we mean up to and including the Cauchy horizon.
In this section, we will use the energy estimates of the previous sections, together with Lemma~\ref{decay},
to obtain a pointwise bound for $\partial_{\tilde r}\phi$, for a fixed $r=r_1>r_-$.
This together with the previously established pointwise bounds for other derivatives of $\phi$, which are valid up to the Cauchy horizon, can then be
used to integrate~\eqref{wave_u} and obtain a pointwise bound $\partial_{\tilde r}\phi$,
up to the Cauchy horizon. Finally, the control of this quantity in $C^1$ will allow us to extend $\phi$ as a classical solution of the wave equation, all the way up to the Cauchy horizon.

\begin{proof}[{\bf Proof of Theorem~\ref{T1}}]
	We proceed in four steps.
	
{\bf (i) Bounding $\partial_{\tilde r}\phi$ for $r=r_1>r_-$.}	
Assume $(\s+1)\kappa_+>\kappa_-$. We now fix $\kappa$ satisfying
$\kappa_-<\kappa<(\s+1)\kappa_+$ and, as before, choose
$\delta>0$ satisfying $\kappa<(\s+1-\delta)\kappa_+$.

We will start by showing that, for a fixed $r=r_1>r_-$, we have
\begin{eqnarray}
\label{pointEstdr}
|\partial_{\tilde r}\phi|(u,r_1,\omega)&\leq& Ce^{-\kappa u}.\label{d}
\end{eqnarray}
Indeed, this follows by the procedure developed in Section~\ref{subSecR0}:  we start
by realizing that for any $\psi=\partial^l_u\Omega^I\phi$, with $l\in\mathbb{Z}_0^+$ and $I$ a mutli-index, we have, in view of~\eqref{first-inequality},
\begin{eqnarray*}
\int_{\Sigma_{r_1}(u_1,u_2)}J^{N_b}_\mu(\psi)n^\mu_{\Sigma_{r_1}}
&\sim&
\int_{u_1}^{u_2}\int_{S^2}\left[(\partial_u\psi)^2+(-D)
(\partial_{\tilde r}\psi)^2+|\slashed\nabla\psi|^2\right](u,r_1,\omega)\,d\omega du,
\\
	&\leq& Ce^{-(2\s+2-\delta)\kappa_+u_1},
\end{eqnarray*}
which implies that
\begin{eqnarray*}
	&&\int_{u_1}^{u_2}\|\partial_u\partial_{\tilde r}\phi(u,r_1,\,\cdot\,)\|^2_{H^2(S^2)}\,du\\
	&&\qquad\qquad\qquad\leq
	C\int_{\Sigma_{r_1}(u_1,u_2)}J^{N_b}_\mu(\partial_u\phi)n^\mu_{\Sigma_{r_1}}\,dV_{\Sigma_{r_1}}\\
	&&\qquad\qquad\qquad\ \ \,+C\sum_{i=1}^3\int_{\Sigma_{r_1}(u_1,u_2)}J^{N_b}_\mu({\it \Omega}_i\partial_u\phi)n^\mu_{\Sigma_{r_1}}\,dV_{\Sigma_{r_1}}\\
	&&\qquad\qquad\qquad\ \ \,+C\sum_{i=1}^3\int_{\Sigma_{r_1}(u_1,u_2)}J^{N_b}_\mu({\it \Omega}_i^2\partial_u\phi)n^\mu_{\Sigma_{r_1}}\,dV_{\Sigma_{r_1}}\\
	&&\qquad\qquad\qquad\leq Ce^{-(2\s+2-\delta)\kappa_+u_1},
\end{eqnarray*}
and allows one to estimate
\begin{eqnarray*}
	&&\sup_{\omega\in\Sd}(\partial_{\tilde r}\phi)^2(u_2,r_1,\omega)\\
	&&\qquad\qquad\leq C
	e^{-(2\s+2-2\delta)\kappa_+u_1}\int_{u_1}^{u_2}e^{(2\s+2-2\delta)\kappa_+u}\|\partial_u\partial_{\tilde r}\phi(u,r_1,\,\cdot\,)\|^2_{H^2(\Sd)}\,du\\
	&&\qquad\qquad\ \ \ \,+
	2\sup_{\omega\in\Sd}(\partial_{\tilde r}\phi)^2(u_1,r_1,\omega)\\
	&&\qquad\qquad\leq C
	e^{-(2\s+2-2\delta)\kappa_+u_1}+2\sup_{\omega\in\Sd}(\partial_{\tilde r}\phi)^2(u_1,r_1,\omega).
\end{eqnarray*}

We can now apply Lemma~\ref{decay}
to $u\mapsto\sup_{\omega\in\Sd}(\partial_{\tilde r}\phi)^2(u,r_1,\omega)$ to finish the proof of~\eqref{pointEstdr}.

{\bf (ii) Bounding $\partial_{\tilde r}\phi$ up to the Cauchy horizon.}
Let
$$u_{r_1}(\uv):=\uv-2r^*(r_1)$$
so that $r_1(u_{r_1}(\uv),\uv)=r_1$.
Integrating the wave equation, in its form~\eqref{wave_u}, along a segment with fixed $\uv$,
from $u_{r_1}(\uv)$ to $\uu<u_{r_1}(\uv)$,
we get
\begin{eqnarray*}
\partial_{\tilde{r}}\phi(\uu,\uv,\omega)&=&
 \partial_{\tilde{r}}\phi(u_{r_1}(\uv),\uv,\omega)
e^{\int_{u_{r_1}(\uv)}^{\uu} G_1(r(\tilde u,\uv))\,d\tilde u}\\ &&-
\int_{u_{r_1}(\uv)}^{\uu}\tilde{S}(\phi)(\tilde u,\uv,\omega)e^{\int_{\tilde{u}}^{\uu} G_1(r(s,\uv))\,ds}\,d\tilde{u}.
\end{eqnarray*}
Choose $0<\eps<\kappa-\kappa_-$ and $r_1$ such that $G_1(r)\geq-\kappa_--\eps$,
for $r_-<r\leq r_1$.
Using~\eqref{d}, and~\eqref{one} and~\eqref{three} to estimate $\tilde{S}(\phi)$,
yields
\begin{eqnarray}
|\partial_{\tilde{r}}\phi(\uu,\uv,\omega)|&\leq&Ce^{-\kappa u_{r_1}(\uv)}
e^{(\kappa_-+\eps)(u_{r_1}(\uv)-\uu)}+\int^{u_{r_1}(\uv)}_{\uu}Ce^{-\kappa\tilde u}e^{(\kappa_-+\eps)(\tilde{u}-\uu)}\,
d\tilde{u}\nonumber\\
&\leq&Ce^{-(\kappa-\kappa_--\eps)u_{r_1}(\uv)}e^{-(\kappa_-+\eps)\uu}+Ce^{-(\kappa_-+\eps) \uu}e^{-(\kappa-\kappa_--\eps)\uu}\nonumber\\
&\leq&C e^{-\kappa\uu},\label{yep}
\end{eqnarray}
for $\uu\in\mathbb{R}$ and $\uv\geq\uu+2r^*(r_1)$.

{\bf (iii) Continuity of $\partial_{\tilde r}\phi$ up to the Cauchy horizon.}
We define $\partial_{\tilde{r}}\phi(\uu,\infty,\omega)$ by
\begin{equation}\label{CH}
\partial_{\tilde{r}}\phi(\uu,\infty,\omega)=
\int^{+\infty}_{\uu}\tilde{S}(\phi)(\tilde u,\infty,\omega)e^{\kappa_-(\tilde{u}-\uu)}\,d\tilde{u}
\end{equation}
Let $-\infty<\overline{u}<\overline{U}<+\infty$. To prove the uniform convergence of $\partial_{\tilde{r}}\phi(\,\cdot\,,\uv,\,\cdot\,)$ to
$\partial_{\tilde{r}}\phi(\,\cdot\,,\infty,\,\cdot\,)$, as $\uv\to\infty$, for
the first variable $\uu$ belonging to $[\overline{u},\overline{U}]$, we write
\begin{eqnarray*}
\partial_{\tilde{r}}\phi(\uu,\uv,\omega)-\partial_{\tilde{r}}\phi(\uu,\infty,\omega)
&=&\partial_{\tilde{r}}\phi(u_{r_1}(\uv),\uv,\omega)
e^{\int_{u_{r_1}(\uv)}^{\uu} G_1(r(\tilde u,\uv))\,d\tilde u}\\
&&+\int^{U}_{\uu}\tilde{S}(\phi)(\tilde u,\uv,\omega)e^{\int_{\tilde{u}}^{\uu} G_1(r(s,\uv))\,ds}\,d\tilde{u}\\
&&-\int^{U}_{\uu}\tilde{S}(\phi)(\tilde u,\infty,\omega)e^{\int_{\tilde{u}}^{\uu} (-\kappa_-)\,ds}\,d\tilde{u}\\
&&+\int^{u_{r_1(\uv)}}_{U}\tilde{S}(\phi)(\tilde u,\uv,\omega)e^{\int_{\tilde{u}}^{\uu} G_1(r(s,\uv))\,ds}\,d\tilde{u}\\
&&-\int^{+\infty}_{U}\tilde{S}(\phi)(\tilde u,\infty,\omega)e^{\int_{\tilde{u}}^{\uu} (-\kappa_-)\,ds}\,d\tilde{u}\\
&=&A+B+C+D+E.
\end{eqnarray*}
Let $\eps>0$.
Again using estimates~\eqref{one} and~\eqref{three} to control~$\tilde{S}$, we can
fix $U$ and $V_1$ sufficiently big so that $|D|+|E|<\frac{\eps}{3}$, for
$\uv\geq V_1$.
Now, using~\eqref{d}, fix $V_2\geq V_1$ sufficiently large so that $|A|<\frac{\eps}{3}$, for $\uv\geq V_2$ and
$\uu\in[\overline{u},\overline{U}]$.
Finally, invoking the uniform convergence, in $[\overline{u},\overline{U}]\times\Sd$, of
$\tilde{S}(\phi)(\,\cdot\,,\uv,\,\cdot\,)$ to $\tilde{S}(\phi)(\,\cdot\,,\infty,\,\cdot\,)$
and of $G_1(r(\uu,\uv))$ to $-\kappa_-$,
as $\uv\to\infty$,
we are allowed to fix $V\geq V_2$ such that $|B+C|<\frac{\eps}{3}$, for $v\geq V$.
For $v\geq V$ and $(\uu,\omega)\in[\overline{u},\overline{U}]\times\Sd$, we have
$$
|\partial_{\tilde{r}}\phi(\uu,\uv,\omega)-\partial_{\tilde{r}}\phi(\uu,\infty,\omega)|<\eps.
$$
This proves the stated uniform convergence. Again, arguing as in~\cite[Proposition~5.2, Step~2]{CGNS3}, $\partial_{\tilde{r}}\phi$
is continuous in ${\cal M}\cup {\cal C}{\cal H}_A^+$.
A simple argument implies that
the derivative of the continuous extension of $\phi$ with respect to $\tilde r$ exists and
coincides with the continuous extension of $\partial_{\tilde r}\phi$.

{\bf (iv) The wave equation is satisfied on the Cauchy horizon.}
To justify that the wave equation~\eqref{wave_u} is satisfied on the Cauchy horizon we
just have to 
differentiate the right-hand side of~\eqref{CH} with respect to~$\uu$. Note that we are not claiming
that $\phi$ is $C^2$ up to the Cauchy horizon but merely that $\partial_{\uu}\partial_{\tilde r}\phi$ exists, is continuous and satisfies~\eqref{wave_u}. We can also
guarantee that $\partial_{\tilde r}\partial_{\uu}\phi$ exists and is continuous. Indeed,
define
$$
\hat{S}=-\,\frac{1}{r}\partial_u-\,\frac{1}{2}\slashed\Delta.
$$
The wave equation can also be written as
\begin{equation}\label{wave_uuu}
\partial_{\tilde r}\partial_{\uu}\phi+\frac{2}{r}\partial_{\uu}\phi=-\hat{S}(\phi).
\end{equation}
This can be integrated to
\begin{equation}\label{integral-form}
\partial_{\uu}\phi(u,r,\omega)=\frac{r_2^2}{r^2}\partial_{\uu}\phi(u,r_2,\omega)
-\int_{r_2}^r\frac{s^2}{r^2}\hat{S}(\phi)(u,s,\omega)\,ds.
\end{equation}
Note that $\partial_{\uu}\phi$ is continuous up to the Cauchy horizon because
it is equal to $\partial_u\phi-\,\frac{D}{2}\partial_{\tilde r}\phi$. Therefore,
equation~\eqref{integral-form} holds with $r_2=r_-$. Another application of the
Fundamental Theorem of Calculus guarantees that
$\partial_{\tilde r}\partial_{\uu}\phi$ exists and is continuous up to the Cauchy horizon
and that~\eqref{wave_uuu} is satisfied also on the Cauchy horizon. Of course,
the fact that $\partial_{\tilde r}\partial_{\uu}\phi$ exists and is continuous up to the Cauchy horizon is enough to guarantee that~\eqref{wave_uuu} is satisfied on the Cauchy horizon.
\end{proof}

\section{Solutions with higher regularity}\label{high}

This section is devoted to the proof of Theorem~\ref{T2}.

\subsection{Wave equations for $\partial_{\tilde r}^n\phi$}
We start by deriving the inhomogeneous wave equations satisfied by higher order $r$-derivatives
of $\phi$.
The commutators of $\Box_g$ and $\partial_r$, and of $\Box_g$ and $\partial_{\tilde r}$, are
$$
[\Box_g,\partial_r]=-D'\partial_r^2+2\Sdois
$$
and
$$
[\Box_g,\partial_{\tilde r}]=-D'\partial_{\tilde r}^2+2\tSdois,
$$
where
$$
\Sdois=\frac{1}{r^2}\partial_v+\left(
\frac{1}{r^2}D-\frac{1}{r}D'-\frac{D''}{2}\right)\partial_r
+\frac{1}{r}\slashed\Delta
$$
and
$$
\tSdois=-\,\frac{1}{r^2}\partial_u+\left(
\frac{1}{r^2}D-\frac{1}{r}D'-\frac{D''}{2}\right)\partial_{\tilde r}
+\frac{1}{r}\slashed\Delta.
$$
Consequently, the function $\partial_r\phi$ satisfies the inhomogeneous wave equation
\begin{eqnarray}
\Box_g(\partial_r\phi)=-D'\partial_r^2\phi+2\Sdois(\phi),\label{wave-one}
\end{eqnarray}
the function $\partial_r^2\phi$ satisfies the inhomogeneous wave equation
\begin{eqnarray*}
\Box_g(\partial_r^2\phi)&=&[\Box_g,\partial_r]\partial_r\phi+\partial_r[\Box_g(\partial_r\phi)]\\
&=&-2D'\partial_r^3\phi-D''\partial_r^2\phi+2\Sdois(\partial_r\phi)+2\partial_r\Sdois(\phi),
\end{eqnarray*}
and, in general, $\partial_r^n\phi$ satisfies the inhomogeneous wave equation
\begin{eqnarray}
\Box_g(\partial_r^n\phi)
&=&-\sum_{l=1}^n\left(n\atop l\right)D^{(l)}\partial_r^{n+2-l}\phi+2\sum_{l=0}^{n-1}\partial_r^l\Sdois(\partial_r^{n-1-l}\phi)\nonumber\\
&=&-nD'\partial_r^{n+1}\phi+2H_n(\phi),\label{wave}
\end{eqnarray}
with
$$
H_n(\phi)=-\,\frac{1}{2}\sum_{l=2}^n\left(n\atop l\right)D^{(l)}\partial_r^{n+2-l}\phi+\sum_{l=0}^{n-1}\partial_r^l\Sdois(\partial_r^{n-1-l}\phi).
$$
We also define
$$
\tilde{H}_n(\phi)=-\,\frac{1}{2}\sum_{l=2}^n\left(n\atop l\right)D^{(l)}\partial_{\tilde r}^{n+2-l}\phi+\sum_{l=0}^{n-1}\partial_{\tilde r}^l\tSdois(\partial_{\tilde r}^{n-1-l}\phi).
$$
Because $T$ ($\tilde{T}$) is a differential operator of order one in $r$ ($\tilde{r}$),
$H_n(\phi)$ ($\tilde{H}_n(\phi)$) involves a sum of derivatives of $\phi$ whose order with respect to $r$ ($\tilde{r}$) is
at most $n$.

Taking into account that the sequence defined by~\eqref{Gn} satisfies
$$
G_1+\frac{n}{2}D'=G_{n+1},
$$
we can write~\eqref{wave} as
$$
\partial_{\uv}\partial_r^{n+1}\phi+G_{n+1}\partial_r^{n+1}\phi=S(\partial_r^n\phi)+H_n(\phi).
$$
Similarly, we have that
\begin{equation}\label{wave_uu}
\partial_{\uu}\partial_{\tilde r}^{n+1}\phi-G_{n+1}\partial_{\tilde r}^{n+1}\phi=
-\tilde{S}(\partial_{\tilde r}^n\phi)-
\tilde{H}_n(\phi).
\end{equation}

\subsection{Derivatives of $\phi$ in $\calS_\s$}
Recall the definition of $\calS_\s$ in~\eqref{Fs}.
\begin{Rmk}\label{dv} Let $l,i_1,i_2,i_3\in\mathbb{Z}_0^+$. Then $\phi\in\calS_\s\ \Rightarrow\ \partial_v^l\Omega_1^{i_1}\Omega_2^{i_2}\Omega_3^{i_3}\phi\in\calS_\s$.
\end{Rmk}
\begin{proof}

For $l=0$ the result follows immediately from the fact that the vector fields $\Omega$ are Killing and tangent to the sphere $\{r=r_0\}\cap \{v=v_0\}$. For the remaining cases it suffices to note that:
\begin{enumerate}[{\rm (i)}]
\item Obviously, $\phi\in\calS_0\ \Rightarrow\ \partial_v^l\phi\in\calS_0$.
\item $\phi\in\calS_1\ \Rightarrow\ \partial_v^l\phi\in\calS_1$.
Recalling that $\partial_{\uv}=\partial_v+\frac{D}{2}\partial_r$, the wave equation~\eqref{wave_v} shows that if $\phi$ vanishes on the event horizon
and $\partial_r\phi(v_0,r_+,\,\cdot\,)=0$, then
$\partial_v\partial_r\phi(v_0,r_+,\,\cdot\,)=0$.
Differentiating both sides of~\eqref{wave_v} once with respect to $v$ we then
conclude that $\partial_v^2\partial_r\phi(v_0,r_+,\,\cdot\,)=0$.
If we keep differentiating both sides of~\eqref{wave_v} with respect to $v$,
we see that
$\partial_v^l\partial_r\phi(v_0,r_+,\,\cdot\,)=0$.
\item $\phi\in\calS_2\ \Rightarrow\ \partial_v^l\phi\in\calS_2$. Before handling the
general case, lets us also go over the case $\s=2$ in detail.
Suppose now that $\phi$ vanishes on the event horizon, $\partial_r\phi(v_0,r_+,\,\cdot\,)=0$
and $\partial_r^2\phi(v_0,r_+,\,\cdot\,)=0$. Using~\eqref{wave-one}
together with the previous paragraph
we conclude that $\partial_v\partial_r^2\phi(v_0,r_+,\,\cdot\,)=0$.
One can also argue that $\partial_v^l\partial_r^2\phi(v_0,r_+,\,\cdot\,)=0$, for all $l>0$.
\item $\phi\in\calS_{n+1}\ \Rightarrow\ \partial_v^l\phi\in\calS_{n+1}$.
More generally, using~\eqref{wave}, if $\phi$ vanishes on the event horizon and its first
$n+1$ derivatives with respect to $r$ vanish at $(v_0,r_+,\,\cdot\,)$,
then
$\partial_v^l\partial_r^m\phi(v_0,r_+,\,\cdot\,)=0$, for all $m\leq n+1$.
\end{enumerate}
\end{proof}

\subsection{Bounding higher $\partial_{\tilde r}$ derivatives of $\phi$}

 The wave equation~\eqref{wave_u} can be used to bound $\partial_{\tilde r}^2\phi$ when $r=r_1>r_-$.
If we integrate the
inhomogeneous wave equation for $\partial_{\tilde r}\phi$ and use the bounds
for derivatives of $\phi$, whose order with respect to $\tilde{r}$ is at most equal to one, then
we can bound $\partial_{\tilde r}^2\phi$ up to the Cauchy horizon if
$2\kappa_-<(s+1)\kappa_+$. The $2\kappa_-$ comes from the fact that
in~\eqref{wave_uu}, with $n=1$, the function in front of $\partial_{\tilde r}^2\phi$ is $G_2$
and $G_2(r_-)=-2\kappa_-$. Moreover, when $l\kappa_-<(s+1)\kappa_+$ we will be able to generalize the previous procedure and establish boundedness of $\partial_{\tilde r}^l\phi$.

\begin{proof}[{\bf Proof of Theorem~\ref{T2}}] We proceed in three steps.

{\bf (i) $\phi\in C^2({\cal M}\cup\cal{CH}_A^+)$.}
Since $\phi\in\calS_\s$ implies $\partial_u\phi=\partial_v\phi\in\calS_\s$,
using~\eqref{yep} applied to $\partial_u\phi$, when $\kappa_-<\kappa<(s+1)\kappa_+$
we have
$$
|\partial_{\tilde r}\partial_u\phi|(u,r,\omega)\leq Ce^{-\kappa u},
$$
for $r_-\leq r\leq r_1$. Then the wave equation~\eqref{wave_u} shows that
on the hypersurface $r=r_1>r_-$ we have that
\begin{equation}\label{estimate-two}
|\partial_{\tilde r}^2\phi|(u,r_1,\omega)\leq Ce^{-\kappa u}.
\end{equation}
Integrating the wave equation~\eqref{wave_uu} with $n=1$ we obtain
\begin{eqnarray*}
\partial_{\tilde{r}}^2\phi(\uu,\uv,\omega)&=&
 \partial_{\tilde{r}}^2\phi(u_{r_1}(\uv),\uv,\omega)
e^{\int_{u_{r_1}(\uv)}^{\uu} G_2(r(\tilde u,\uv))\,d\tilde u}\\ &&-
\int_{u_{r_1}(\uv)}^{\uu}(\tilde{S}(\partial_{\tilde{r}}\phi)+\tilde{H}_1(\phi))(\tilde u,\uv,\omega)e^{\int_{\tilde{u}}^{\uu} G_2(r(s,\uv))\,ds}\,d\tilde{u}.
\end{eqnarray*}
The derivatives of $\phi$ appearing inside the integral
have order at most one with respect to $\tilde r$, namely they are
$\partial_u\phi$, $\slashed\Delta\phi$, $\partial_{\tilde{r}}\phi$, $\partial_{\tilde{r}}\partial_u\phi$ and
$\partial_{\tilde{r}}\slashed\Delta\phi$.
Moreover, $$|\partial_{\tilde{r}}\slashed\Delta\phi|(u,r,\omega)\leq Ce^{-\kappa u},$$
for $r_-\leq r\leq r_1$. Therefore, under the assumption that $2\kappa_-<\kappa<(s+1)\kappa_+$
we obtain that
$$
|\partial_{\tilde r}^2\phi|(u,r,\omega)\leq Ce^{-\kappa u},
$$
for $r_-\leq r\leq r_1$. As, in addition, when $\kappa_-<\kappa<(s+1)\kappa_+$, $\partial_u^2\phi$,
$\partial_{\tilde r}{\it \Omega}\phi$ and $\partial_u{\it\Omega}\phi$
are continuous on
${\cal M}\cup\cal{CH}_A^+$
we have that $\phi\in C^2({\cal M}\cup\cal{CH}_A^+)$ when $2\kappa_-<(s+1)\kappa_+$.

{\bf (ii) $\phi\in C^3({\cal M}\cup\cal{CH}_A^+)$.}
Let us consider another specific case, $n=2$, before analyzing the general situation: repeating the previous argument, using~\eqref{estimate-two} applied to $\partial_u\phi$, when
$2\kappa_-<\kappa<(s+1)\kappa_+$ we have
$$
|\partial_{\tilde r}^2\partial_u\phi|(u,r,\omega)\leq Ce^{-\kappa u},
$$
for $r_-\leq r\leq r_1$. The wave equation~\eqref{wave-one} shows that
on the hypersurface $r=r_1>r_-$ we have that
$$
|\partial_{\tilde r}^3\phi|(u,r_1,\omega)\leq Ce^{-\kappa u}.
$$
Since the above mentioned derivatives, $\partial_{\tilde r}^2\partial_u\phi$ and
$\partial_{\tilde r}^2\slashed\Delta\phi$ are controlled when $2\kappa_-<\kappa<(s+1)\kappa_+$,
integrating the wave equation~\eqref{wave_uu} with $n=2$, when $3\kappa_-<\kappa<(s+1)\kappa_+$ we obtain
that
$$
|\partial_{\tilde r}^3\phi|(u,r,\omega)\leq Ce^{-\kappa u},
$$
for $r_-\leq r\leq r_1$. All other third order derivatives of $\phi$
are continuous when $2\kappa_-<\kappa(s+1)\kappa_+$ and we are able to conclude that $\phi\in C^3({\cal M}\cup\cal{CH}_A^+)$ when $3\kappa_-<(s+1)\kappa_+$.

{\bf (iii) $\phi\in C^l({\cal M}\cup\cal{CH}_A^+)$.}
The general case now clearly follows by induction: in fact, when $l\kappa_-<\kappa<(s+1)\kappa_+$ we obtain that
$$
|\partial_{\tilde r}^l\phi|(u,r,\omega)\leq Ce^{-\kappa u},
$$
for $r_-\leq r\leq r_1$. In conclusion, $\phi\in C^l({\cal M}\cup\cal{CH}_A^+)$ provided $l\kappa_-<(s+1)\kappa_+$.
\end{proof}

\section{Blow up in $C^1$ and in $H^1$}\label{ch}

\subsection{Blow up in $C^1$}
\label{subSecBlowC1} This subsection is devoted to the proof of Theorem~\ref{C1-blow-up}. Let us start by sketching the main ideas for this proof.
Suppose that $\phi$ is spherically symmetric.
If for all large $\uu$ we have that $-\partial_{\uu}\phi(\uu,v_0,\,\cdot\,)$ is positive, then it turns out that
$-\partial_{\uu}\phi(\uu,\uv,\,\cdot\,)$ and
$\partial_{\uv}\phi(\uu,\uv,\,\cdot\,)$ are positive for $(\uu,\uv)\in[u_{r_0}(v_0),\infty[\times[v_0,\infty[$.
This fact can be used to propagate a lower bound for $-\partial_{\uu}\phi$, at $v=v_0$,
all the way up to the Cauchy horizon. We can then
obtain a lower bound for $\partial_{\uv}\phi$ and a negative upper bound for
$\partial_{\tilde r}\phi$, 
which can be used to obtain the desired blow up result.

\begin{proof}[{\bf Proof of Theorem~\ref{C1-blow-up}}] We proceed in six steps.

{\bf (i) Initial data for $-\partial_{\uu}\phi$.}
Assume first that $\phi$ is spherically symmetric and that
the restriction of $\phi$ to the ingoing null hypersurface $v=v_0$, through the event horizon,
vanishes to order $s$ and does not vanish to order $s+1$,
on the event horizon.
Then there exist constants $0<c<C$ such that (eventually replacing $\phi$ by $-\phi$)
$$
c(r_+-r)^s\leq-\partial_r\phi(v_0,r,\,\cdot\,)\leq C(r_+-r)^s,
$$
for $r_0\leq r\leq r_+$.
As $\partial_{\uu}\phi=-\,\frac{D}{2}\partial_r\phi$, using~\eqref{Destimate} we have
$$
c(r_+-r(\uu,v_0))^{s+1}\leq-\partial_{\uu}\phi(\uu,v_0,\,\cdot\,)\leq C(r_+-r(\uu,v_0))^{s+1},
$$
for $\uu\geq u_{r_0}(v_0)$. 
According to~\eqref{r-star} and~\eqref{r-star-vu}, for $r_0\leq r\leq r_+$,
there exist constants $0<c<C$ such that
$$
ce^{\kappa_+(\uv-\uu)}\leq r_+-r(\uu,\uv)\leq Ce^{\kappa_+(\uv-\uu)}.
$$
Thus, we get
$$
ce^{-(s+1)\kappa_+\uu}\leq-\partial_{\uu}\phi(\uu,v_0,\,\cdot\,)\leq Ce^{-(s+1)\kappa_+\uu},
$$
for $\uu\geq u_{r_0}(v_0)$.

{\bf (ii) $-\partial_{\uu}\phi$ in $\cal M$.}
Since $\phi$ is spherically symmetric
the wave equation reduces to
$$
\partial_{\uv}(r\partial_{\uu}\phi)=-\partial_{\uu}r\,\partial_{\uv}\phi.
$$
According to~\cite[Lemma~B.1]{CGNS3},
$-\partial_{\uu}\phi(\uu,\uv,\,\cdot\,)$ and
$\partial_{\uv}\phi(\uu,\uv,\,\cdot\,)$ are positive for
$(\uu,\uv)\in[u_{r_0}(v_0),\infty[\times[v_0,\infty[$.
So $\uv\mapsto -r\partial_{\uu}\phi(\,\cdot\,,\uv,\,\cdot\,)$ is an increasing
function. This implies that
$$
ce^{-(s+1)\kappa_+\uu}\leq-\partial_{\uu}\phi(\uu,\uv,\,\cdot\,),
$$
for $(\uu,\uv)\in[u_{r_0}(v_0),\infty[\times[v_0,\infty[$.

{\bf (iii) $-\partial_{\uv}\phi$ for $r_-\leq r\leq r_0$.}
Now we integrate the following (spherically symmetric) version of the wave equation
$$
\partial_{\uu}(r\partial_{\uv}\phi)=-\partial_{\uv}r\,\partial_{\uu}\phi
$$
between $+\infty$ and $\uu>u_{r_0}(\uv)$.
Taking into account that in this region
$$
-\partial_{\uv}r(\uu,\uv)=-\,\frac{D}{2}(r(\uu,\uv))\geq
c(r_+-r(\uu,\uv))\geq ce^{\kappa_+(\uv-\uu)},
$$
we obtain
\begin{eqnarray*}
(r\partial_{\uv}\phi)(\uu,\uv,\,\cdot\,)&=&(
r\partial_{\uv}\phi)(+\infty,\uv,\,\cdot\,)+\int_{+\infty}^{\uu}
(-\partial_{\uv}r\,\partial_{\uu}\phi)(\tilde{u},\uv,\,\cdot\,)\,
d\tilde{u}\\
&=&\int^{+\infty}_{\uu}[
(-\partial_{\uv}r)\,(-\partial_{\uu}\phi)](\tilde{u},\uv,\,\cdot\,)\,
d\tilde{u}\\
&\geq&\int^{+\infty}_{\uu}ce^{\kappa_+(\uv-\tilde{u})}
e^{-(s+1)\kappa_+\tilde{u}}\,d\tilde{u}\\
&=&ce^{\kappa_+(\uv-(s+2)\uu)}.
\end{eqnarray*}
On the hypersurface $r=r_0$ we have $\uu=\uv-2r^*(r_0)$, and so
$$
\partial_{\uv}\phi(u_{r_0}(\uv),\uv,\,\cdot\,)\geq
ce^{-(s+1)\kappa_+\uv}.
$$
As $-\partial_{\uu}(r\partial_{\uv}\phi)$ is positive, it follows that
$$
\partial_{\uv}\phi(\uu,\uv,\,\cdot\,)\geq
ce^{-(s+1)\kappa_+\uv},
$$
for $u_{r_0}(v_0)\leq\uu\leq u_{r_0}(\uv)$.

{\bf (iv) $\partial_{\tilde r}\phi$ for $r_-\leq r\leq r_0$.}
In the region $u_{r_0}(v_0)\leq\uu\leq u_{r_0}(\uv)$,
according to~\eqref{r-star} and~\eqref{r-star-vu},
there exist constants $0<c<C$ such that
$$
ce^{-\kappa_-(\uv-\uu)}\leq r(\uu,\uv)-r_-\leq Ce^{-\kappa_-(\uv-\uu)}.
$$
Therefore,
\begin{eqnarray}
\partial_{\tilde r}\phi(\uu,\uv,\,\cdot\,)&=&
\left(\frac{2}{D}\partial_{\uv}\phi\right)(\uu,\uv,\,\cdot\,)\leq
-ce^{\kappa_-(\uv-\uu)}e^{-(s+1)\kappa_+v}\nonumber\\
&=&-ce^{-\kappa_-\uu}e^{(\kappa_--(s+1)\kappa_+)\uv}.\label{lower}
\end{eqnarray}

{\bf (v) Blow up.}
For $\kappa_->(s+1)\kappa_+$ the right-hand side of~\eqref{lower} goes to $-\infty$ as
$v$ goes to $+\infty$. In this case $\phi$ does not extend to a $C^1$
function up to the Cauchy horizon.

Suppose now that $\phi$ is not spherically symmetric. Its spherically
mean
\begin{equation}\label{mean}
\psi(\uu,\uv):=\frac{1}{4\pi}\int_{\Sd}\phi(\uu,\uv,\omega)\,d\omega
\end{equation}
is also a solution of the wave equation.
According to our hypotheses, $\psi$ is not $C^1$ up to the Cauchy horizon.
Therefore $\phi$ cannot be $C^1$ up to the Cauchy horizon.

{\bf (vi) Uniform blow up.} The previous analysis only provides blow up of the $C^1$ norm of $\phi$ along null rays
$\uu=u_1$, with $u_1\geq u_{r_0}(v_0)$. To extend the result to all outgoing null rays  intersecting $\cal{CH}_A^+$,  assume that $\partial_{\tilde r}\phi$ is bounded along some $u=u_2<u_{r_0}(v_0)$. Then we would be able to solve the spherically symmetric wave equation sideways, with characteristic initial data provided by the spherical mean $\psi$ along
$[u_2,u_1]\times \{v_0\}\cup \{u_2\}\times[v_0,\infty]$, for $u_1\geq u_{r_0}(v_0)$. But since the spacetimes region  $[u_2,u_1]\times[v_0,\infty]$ is compact, local well posedness (see for instance~\cite[Theorem 4.5]{CGNS1}) and the regularity of the initial data would imply boundedness of $\partial_{\tilde r}\psi$ along $u=u_1$. This is a contradiction.
\end{proof}

\subsection{Blow up in $H^1$}
To prove that $\phi$ does not belong to $H^1$ it is enough to prove that its
spherically symmetric part does not belong to $H^1$. Moreover, the negative upper bound~\eqref{lower}
applied to the spherical mean can be used to obtain a lower bound for the $H^1$ norm of $\phi$.

\begin{proof}[{\bf Proof of Theorem~\ref{H1-blow-up}}] We proceed in three steps.

{\bf (i) The $H^1$ norm of $\phi$.}
To define a $H^1$ norm on $\cal M\cup\cal{CH}_A^+$, we define a Riemannian metric $h$ on $\cal M\cup\cal{CH}_A^+$
in the usual way. After choosing a unit timelike vector field $X$, we
let
$$
h_{\mu\nu}=g_{\mu\nu}+2X_\mu X_\nu.
$$
Our choice of $X$ is
$$
X=\frac{1}{\sqrt{2}}(-\partial_{\uu}-\partial_{\tilde r})
=\frac{1}{\sqrt{2}}\left(-\partial_u-\left(1-\,\frac{D}{2}\right)\partial_{\tilde r}\right).
$$	
This leads to
$$
h=\left(1+\frac{D^2}{4}\right)\,du^2+D\,dud\tilde{r}+d\tilde{r}^2+r^2\sigma_{\mathbb{S}^2}.
$$
The square of the norm of the gradient of $\phi$ is
$$
h(\nabla\phi,\nabla\phi)=(\partial_{\uu}\phi)^2+(\partial_{\tilde r}\phi)^2
+|\slashed\nabla\phi|^2.
$$
The volume element on ${\cal M}\cup\cal{CH}_A^+$ is
$$
dV_{\cal M}=r^2\,d\omega d\tilde{r}du.
$$	
Now we may define the $H^1$ norm of $\phi$ to be
\begin{eqnarray*}
	\|\phi\|_{H^1(\cal{M})}^2&=&\|\phi\|_{L^2(\cal{M})}^2+
	\|\nabla\phi\|_{L^2(\cal{M})}^2\\
	&=&
	\int_{\cal{M}}\phi^2\,dV_{\cal M}+\int_{\cal{M}}h(\nabla\phi,\nabla\phi)\,dV_{\cal M}.
\end{eqnarray*}

{\bf (ii) Decomposition of the $H^1$ norm.}
Again, let $\psi$ be the spherical mean of $\phi$. We remark that
\begin{equation}\label{squares}
\|\nabla\phi\|_{L^2(\cal{M})}^2=\|\nabla(\phi-\psi)\|_{L^2(\cal{M})}^2+\|\nabla\psi\|_{L^2(\cal{M})}^2.
\end{equation}
Indeed, this follows from
\begin{eqnarray*}
&&	\int_{\cal{M}}h(\nabla\phi,\nabla\psi)\,dV_{\cal M}\\
	&&\qquad=\int_{\cal{M}}
	\left((\partial_{\uu}\phi)(\partial_{\uu}\psi)+
	(\partial_{\tilde{r}}\phi)(\partial_{\tilde{r}}\psi)
	\right)r^2\,d\omega d\tilde{r}du\\
	&&\qquad=\frac{1}{4\pi}\int_{\cal{M}}
	\left[(\partial_{\uu}\phi)
	\left(\int_{\Sd}\partial_{\uu}\phi\,d\omega\right)+
	(\partial_{\tilde r}\phi)
	\left(\int_{\Sd}\partial_{\tilde r}\phi\,d\omega\right)
	\right]r^2\,d\omega d\tilde{r}du\\
		&&\qquad=\int_{\cal{M}/SO(3)}
	\left(\left(\int_{\Sd}\partial_{\uu}\phi\,d\omega\right)^2+
	\left(\int_{\Sd}\partial_{\tilde r}\phi\,d\omega\right)^2
	\right)r^2\,d\tilde{r}du\\
	&&\qquad=\int_{\cal{M}}h(\nabla\psi,\nabla\psi)\,dV_{\cal M}.
\end{eqnarray*}
For the second equality we used~\eqref{mean} and the fact that $\phi$ belongs to
$C^\infty(\cal{M})$.

{\bf (iii) Blow up.}
From~\eqref{squares}, to prove that $\phi$ does not belong to $H^1_{{\rm loc}}$ it is enough to prove
that $\psi$ does not belong to $H^1_{{\rm loc}}$. Using~\eqref{lower},
the $L^2$ norm of $\partial_{\tilde r}\psi$ is bounded below by
\begin{eqnarray*}
&&\int_{u_0}^{u_1}\int_{r_-}^{r_2}\int_{\Sd}(\partial_{\tilde r}\psi)^2
(u,r,\omega)\,dV_{\cal M}\\
&&\qquad=\int_{u_0}^{u_1}\int_{r_-}^{r_2}\int_{\Sd}(\partial_{\tilde r}\psi)^2
(u,r,\omega)r^2\,d\omega d\tilde{r}du\\
&&\qquad=\frac{1}{2}\int_{u_0}^{u_1}\int_{v_{r_2}(u)}^{+\infty}\int_{\Sd}(\partial_{\tilde r}\psi)^2
(u,v,\omega)(-D(r(u,v)))r^2(u,v)\,d\omega d\uv du\\
&&\qquad\geq c
\int_{u_0}^{u_1}\int_{v_{r_2}(u)}^{+\infty}\int_{\Sd}
e^{-2\kappa_-\uu}e^{2(\kappa_--(s+1)\kappa_+)\uv}e^{-\kappa_-(\uv-\uu)}\,
d\omega d\uv d\uu\\
&&\qquad=
\int_{u_0}^{u_1}\int_{v_{r_2}(u)}^{+\infty}\int_{\Sd}e^{-\kappa_-\uu}
e^{(\kappa_--2(s+1)\kappa_+)\uv}\,
d\omega d\uv d\uu,
\end{eqnarray*}
where $r_2>r_-$ and $u_0<u_1<u_{r_0}(v_1)$. This integral is infinite provided
$\kappa_->2(s+1)\kappa_+$. The extension of the $H^1$ blow up to any neighborhood of $\cal{CH}_A^+$ follows once again by local well posedness, as in the end of Subsection~\ref{subSecBlowC1}.
\end{proof}

\section*{Acknowledgements}

We thank J.\ Natário and J.D.\ Silva for useful comments on a preliminary version of this paper.
We also thank Anne Franzen for sharing and allowing us to use Figure~\ref{figura}.
This work was partially supported by FCT/Portugal through UID/MAT/04459/2013 and grant (GPSEinstein) PTDC/MAT-\-ANA/1275/2014.

\end{document}